\documentclass[3p,times,procedia]{elsarticle}
\usepackage{nupha_ecrc}
\usepackage{amsmath}
\usepackage{graphicx}

\volume{00}

\firstpage{1}

\journalname{Nuclear Physics A}

\runauth{M. Luzum}

\jid{nupha}

\jnltitlelogo{Nuclear Physics A}

\usepackage{amssymb}

\usepackage[figuresright]{rotating}

\begin{document}

\begin{frontmatter}



\dochead{XXVIIIth International Conference on Ultrarelativistic Nucleus-Nucleus Collisions\\ (Quark Matter 2019)}

\title{
System response to the initial energy-momentum tensor in relativistic heavy-ion collisions
}


\author[USP]{Jefferson Sousa}
\address[USP]
{Institute of Physics, University of Sao Paulo, SP, Brazil}
\author[UIUC]{Jorge Noronha}
\address[UIUC]{%
	Department of Physics, University of Illinois at Urbana-Champaign,
	Urbana, IL, USA
}
\author[USP]{Matthew Luzum}

%

\begin{abstract}
The evolution of a relativistic heavy-ion collision is typically understood as a process that transmutes the initial geometry of the system into the final momentum distribution of observed hadrons, which can be described via a cumulant expansion of the initial distribution of energy density and is represented at leading order as the well-known eccentricity scaling of anisotropic flow.
We summarize a proposed extension of this framework to include the contribution from initial momentum-space properties, as encoded in other components of the energy-momentum tensor.  Numerical tests validate  this proposal.

%
\end{abstract}


\end{frontmatter}


\section{Introduction}
In relativistic heavy-ion collisions, distinct azimuthal dependences are observed in  measured particle correlations.  These observed anisotropies are typically understood as arising as a response to asymmetries in the initial geometry of  the system.  Despite the complicated nature of the system evolution, this understanding can be codified in simple relations such as the famous eccentricity scaling of elliptic flow
$V_2 = \kappa_2 \mathcal E_2$.

These relations are quite powerful --- in separating the effects of the early-time evolution of the system from the properties of the QGP medium in a simple way, one can devise observables that isolate each from the other, gain an understanding of exactly which properties of the system determine various observables and which are irrelevant, and eliminate the need for computationally expensive simulations in many cases.

Until now, only the system response to the initial geometry (via the distribution of energy) has been characterized in this way.  However, it is expected that other degrees of freedom --- such as the momentum and stress as encoded in other components of the initial energy-momentum tensor $T^{\mu\nu}$ --- should also play a role.  It is interesting to similarly characterize the system response to these properties of the initial state.  That way, one can identify more precisely which properties of the early-time system contribute to the final flows and potentially constrain their values from experimental measurements.  Given that these properties are not necessarily well understood from first principles, and that they are expected to play an increasingly important role in small collision systems, this may have significant importance for future studies.

\section{Cumulant Expansion}
\label{cumulants}

As a reminder, one can understand eccentricity scaling as the first term in a systematic expansion~\cite{Gardim:2011xv}, devised using simple ansatze.  
The goal is to find a relation involving the state of the system at an early time and the final distribution of particles.   The most success has been achieved in describing the azimuthal anisotropy of this distribution, as characterized by Fourier coefficients in azimuthal angle $dN/d\phi = N \sum V_n e^{in\phi}$.
%

Each coefficient $V_n$ is a vector in the transverse plane (here represented as a complex number), representing azimuthal anisotropy in a particular rotational mode labeled by $n$ in a particular event.
%

First, we assume that the final observable in question (for example $V_n$) is accurately determined only by the energy-momentum tensor at some early time $T^{\mu\nu}(\tau = \tau_0, \vec x)$\footnote{It is also likely to depend on conserved currents of the system $j^\mu(\tau_0,\vec x)$.  This will be addressed in a future publication.}.  Then we postulate a hierarchy of scales, such that the structure of this initial condition at larger length scales has a stronger effect on a given final observable than structure at smaller length scales.  
%

The usual cumulant expansion~\cite{Teaney:2010vd} can be derived under the assumption that the only relevant component of $T^{\mu\nu}$ is the energy density --- that is, only the initial geometry of the system matters.  One decomposes the energy density into cumulants, which are naturally ordered in terms of  relevant length scales.   Explicitly, one considers a Fourier transform of  the energy density $\rho = T^{\tau\tau}$
\begin{align}
e^{W(\vec k)} &\equiv \int d^2 x\ e^{-i\vec k\cdot \vec x}\rho(\vec x)
\end{align}
which defines a cumulant generating function $W(\vec k)$.  Small values of $k \equiv |\vec k|$ represent large length scales.  $W$ can therefore be decomposed into a Taylor series around $k=0$ (and separated into well-defined rotational modes)
\begin{align}
W(\vec k) &= \sum_{n=-\infty}^{\infty} \sum_{m=|n|}^{\infty} W_{n,m} k^m e^{-in\phi_k},
\end{align}
where $\phi_k$ is the angle of  the Fourier variable $\vec k$.  The index $n$ labels the rotation mode, while $m$ represents the order in length scale --- large scales are represented by the small $k$ behavior of $W$, which are represented by cumulants $W_{n,m}$ with small $m$.  The most important property at each harmonic $n$ is represented by  the lowest  cumulant $W_{n,n}$, with higher values of $m$ representing subleading  behavior. 

Assuming $\rho(\vec x)$ is sufficiently well behaved that the infinite set of cumulants contains all the same information,  our first ansatz says that $V_n$ is some function of these  cumulants $V_n = V_n(\{W_{n',m'}\})$.  Our second ansatz can be written as
\begin{align}
m > m' \implies \frac{\partial V_n}{\partial {W_{n',m}}}
<  \frac{\partial V_n}{\partial {W_{n'',m'}}}.
\end{align}
That is, the system response to the initial condition is stronger for larger $m$, and weaker as $m$ becomes larger.

Next we assume that the functional dependence can be captured via a power series.  
\begin{align}
\label{leadingorder}
V_{n} \simeq &\sum_{m=n}^{m_{max}} \kappa_{n,m} W_{n,m}  
 + \sum_{l=1}^{m_{max}} \sum_{m=l}^{m_{max}}\sum_{m' = |n-l|}^{m_{max}} \kappa_{l,m,m'}  W_{l,m} W_{n-l,m'}   + O(W^3) .
\end{align}
For $n = 1, 2, 3$, for example, the leading term is the lowest cumulant $W_{n,n}$.  

Finally, since $V_n$ is dimensionless, it is customary to divide each cumulant by a scale to make dimensionless ratios, so that each response coefficient $\kappa$ is also dimensionless.   A natural scale is the system radius $R$ defined by
%
%
%
%
%
\begin{align}
R &= \sqrt{\langle r^2 \rangle_\epsilon - \left|\langle re^{i\phi} \rangle_\epsilon\right|^2 } \equiv \sqrt{ \frac{\int d^2x\ r^2  T^{\tau\tau}(\bf x)}{\int d^2x\  T^{\tau\tau}(\bf x)} - \left|
	\frac{\int d^2x\ r e^{i\phi}\  T^{\tau\tau}(\bf x)}{\int d^2x\  T^{\tau\tau}(\bf x)}
	\right|^2}
\end{align}
so we can define dimensionless eccentricities  
$\mathcal E_{n,m}\equiv -W_{n,m}/R^m$.
%
\section{Adding other $T^{\mu\nu}$ components }
Our proposal is to include contributions  additively  to the energy density before performing the cumulant expansion.  Explicitly, we start with
\begin{align}
\rho(\vec x) &= T^{\tau\tau} (\vec x) - \alpha \partial_i T^{\tau i}(\vec x) + \beta \partial_i \partial_j T^{ij}(\vec x),
\end{align}
with all relations from
Sec.~\ref{cumulants}
remaining valid.   As an explicit example, the lowest order estimators for harmonics $n = 2,3$ are then
\begin{align}
V_2^{(\rm{est})} &= \kappa_2\mathcal E_2(\alpha,\beta) \equiv - \frac{2\kappa_2}{R^2}W_{2,2}(\alpha,\beta)\\
&= -\frac{\kappa_2}{R^2} \left[
\langle r^2e^{i2\phi}\rangle_\epsilon - 2\alpha\langle re^{i\phi}\rangle_u - 4\beta\langle 1\rangle_c 
- \left(\langle re^{i\phi}\rangle_\epsilon - \alpha\langle 1\rangle_u\right)^2
\right],\label{v2est}\\
V_3^{(\rm{est})} &= \kappa_3\mathcal E_3(\alpha,\beta) \equiv - \frac{6\kappa_3}{R^3}W_{3,3}(\alpha,\beta)\\
&= -\frac{\kappa_3}{R^3} \left[
\langle r^3e^{i3\phi}\rangle_\epsilon - 3\alpha\langle r^2e^{i2\phi}\rangle_u - 12\beta\langle re^{i\phi}\rangle_c - 2\left(\langle re^{i\phi}\rangle_\epsilon - \alpha\langle 1\rangle_u\right)^3
\right],\label{v3est}
\end{align}
where the brackets are defined as
\begin{align}
\langle \ldots \rangle_\epsilon &= \frac{\int d^2x \ldots T^{\tau\tau}(\vec x)}{\int d^2x  T^{\tau\tau}(\vec x)};\qquad
\langle \ldots \rangle_u = \frac{\int d^2x \ldots U(\vec x)}{\int d^2x  T^{\tau\tau}(\vec x)};\qquad
\langle \ldots \rangle_c = \frac{\int d^2x \ldots C(\vec x)}{\int d^2x  T^{\tau\tau}(\vec x)}
\end{align}
with
\begin{align}
U(\vec x) &\equiv T^{\tau x} + i T^{\tau y};\qquad
C(\vec{x}) \equiv \frac{1}{2}\left(T^{x x}-T^{y y}\right)+i T^{x y}.
\end{align}

The coefficients $\alpha$ and $\beta$ represent the relative importance of the initial momentum density and stress, respectively, for determining final observables.  They thus represent information about the system response to the initial conditions, similar to the response coefficients $\kappa$.    However, while there is a different coefficient $\kappa$ for every term in the series \eqref{leadingorder}, there are only two additional coefficients introduced here.  Thus our proposal is very restrictive, and we can use this fact to validate it.  If our framework, with a single value of $\alpha$ and a single value of  $\beta$, captures the effect of the system evolution, then we have strong evidence that it is correct. Note in particular that the same coefficients appear in the estimators for $V_2$ and $V_3$.

In the limit $(\alpha,\beta)\to (0,0)$, the effects of initial momentum and stress are neglected, and the estimators revert to the usual definitions,\footnote{Note that in the case of $m>2$, our denominator is slightly different than the traditional choice.} capturing effects only from the initial energy density.

%

\section{Numerical validation and conclusions}

Various tests were performed to validate the proposal --- i.e., that the resulting estimators accurately predict the final flow on an event-by-event basis.    Here we present one such test, with a state-of-the art simulation utilizing initial conditions from IP-Glasma~\cite{Schenke:2012wb, Schenke:2012fw}, viscous hydrodynamic evolution with transport coefficients from a recent Bayesian analysis~\cite{Bernhard:2018hnz}, and UrQMD~\cite{Bass:1998ca, Bleicher:1999xi} as a hadron cascade afterburner.  Here we show results for $p_T$-integrated elliptic and triangular flow. More details can be found in  \cite{prep}.

These calculations involve many simulations, with fluctuating initial conditions.  We quantify the quality of the proposed estimator  using the vector version of  the Pearson Correlation coefficient between the flow  $V_n$ in each event and the appropriate estimator \eqref{v2est} or \eqref{v3est}: 

\begin{align}
Q_n &= \frac
{\left\langle V_n^{(\rm{est})} V_n^*\right\rangle}
{\sqrt{\left\langle \left| V_n^{(\rm{est})} \right|^2 \right\rangle \left\langle \left| V_n \right|^2 \right\rangle}}.
\end{align}

This quantifies the linear correlation, such that a value of 1 signifies a perfect estimator in every event $V_n = V_n^{(\rm{est})}$, while a value of 0 indicates no linear correlation.

\begin{figure}
\begin{center}
\includegraphics[width=\linewidth]{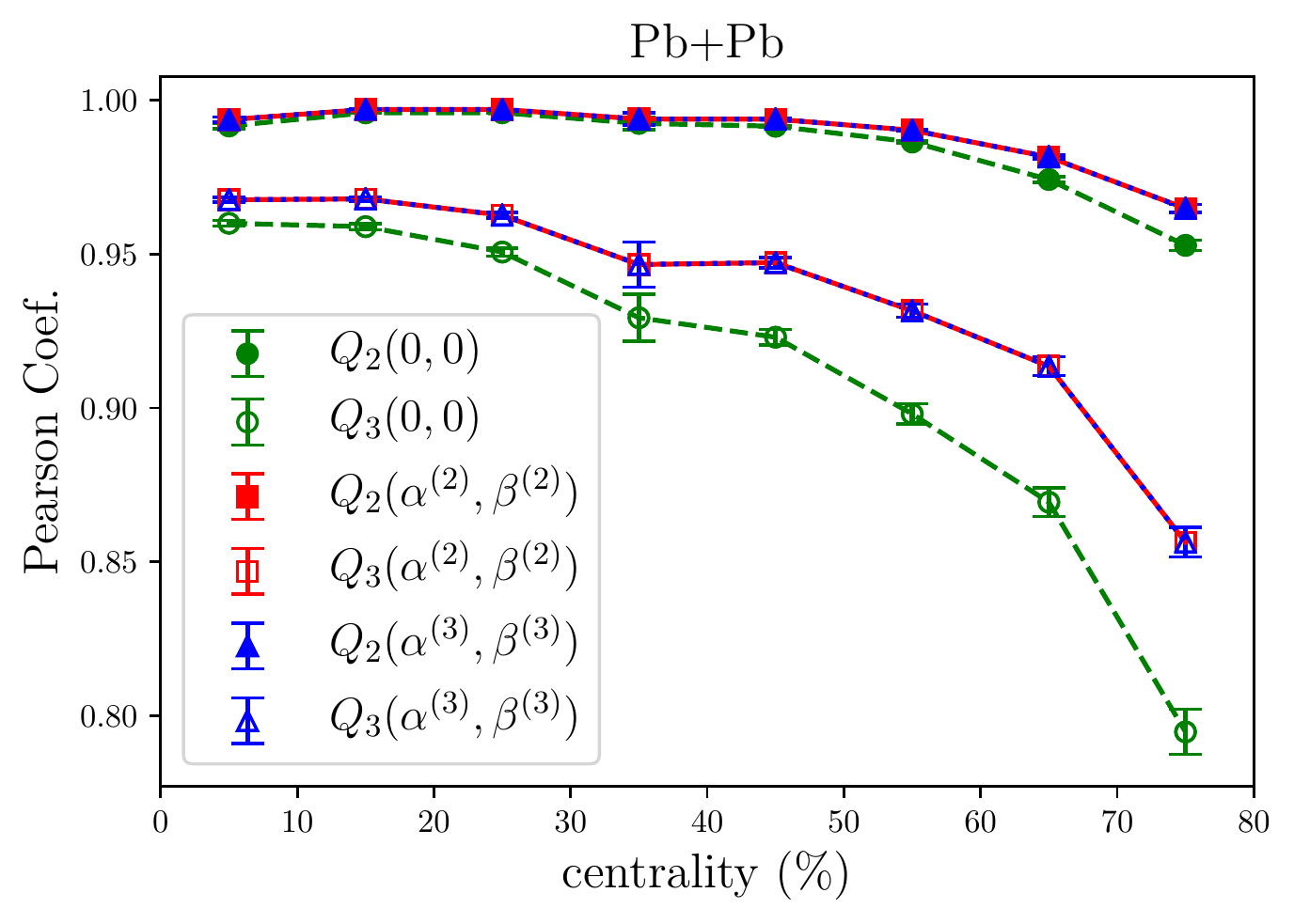}
\end{center}
\caption{\label{Pearson}Linear (Pearson)  correlation coefficient $Q_n(\alpha,\beta)$ between the final flow coefficient $V_n$  and a predictor that includes the effects of initial momentum and stress.   $Q_n(0,0)$ with green dashed connecting lines represents the result for a predictor constructed only from the energy density.  The red solid lines represent $Q_2$ and $Q_3$ with values of $\alpha$ and $\beta$ chosen to maximize the correlation coefficient for $n=2$ ($\alpha^{(2)}, \beta^{(2)}$) and $n=3$ ($\alpha^{(3)}, \beta^{(3)}$) respectively.  The blue dotted line represents the Pearson coefficient with values of $\alpha$ and $\beta$ that maximize the opposite harmonic, showing that $n=2$  and $n=3$  do not require different values of these response coefficients and they are truly independent of harmonic, as predicted from this framework.}
\end{figure}

In Fig.~\ref{Pearson} we can see that, while the traditional eccentricities (obtained by setting $\alpha=\beta=0$) already are very good estimators, adding the effects of initial momentum and stress improve the estimators.  The red solid lines represent the results when $\alpha$ and $\beta$ are varied  so that the best estimator is obtained, independently for each harmonic.  The blue dotted lines represent the result when the best value of the response coefficients for $n=2$ is used for the estimator of $V_3$, and vice versa.  

This is striking validation for our framework --- not only does it more accurately capture the effect of the initial conditions, but the very non-trivial prediction of a single value of response coefficients $\alpha$ and $\beta$, regardless of harmonic, is verified.

With this new understanding, it is now possible to identify more precisely which aspects of the early-time system have an affect on final observables, and potentially to better separate and constrain various physical effects.  While these new degrees of freedom are subdominant in large collision systems, they may be more important in small systems.

\medskip
{\it Acknowledgements.}
%
J.S.~was supported by a CAPES fellowship. 
M.L.~was supported by FAPESP grants 2016/24029-6, 2017/05685-2, and 2018/24720-6.  








\end{document}